\documentclass{PoS}

\title{On the conformal anomaly of k-strings}

\ShortTitle{On the conformal anomaly of k-strings}

\author{\speaker{Pietro Giudice}, Ferdinando Gliozzi, Stefano Lottini\\
Dipartimento di Fisica Teorica, Universit\`a di Torino and INFN, 
sezione di Torino, Italy \\
E-mail: \email{giudice,gliozzi,lottini@to.infn.it}}

\abstract{
We discuss the long distance behaviour of the flux tube associated to baryon 
vertices and argue that, if the gauge system admits stable k-strings, 
 the conformal field theory describing this string in the IR 
has conformal anomaly $c=(d-2)\sigma_k/\sigma$, 
where $\sigma_k$ is the k-string  tension and $\sigma$ that of the
fundamental representation.  We check this result in a 3D $\Z_4$ gauge 
model at finite temperature, where a string effect directly related to 
$c$ can be clearly identified.}

\FullConference{The XXV International Symposium on Lattice Field Theory\\
		 July 30 - August 4 2007\\
		 Regensburg, Germany}

\usepackage{mathrsfs}
\usepackage{epsfig}

\newcommand{\Z}{\mathbb{Z}}

\newcommand{\R}{\mathscr{R}}

\newcommand{\bra}{\langle}
\newcommand{\ket}{\rangle}

\newcommand{\sun}{\mathop{\rm SU}(N)}

\newcommand{\eq}{\begin{equation}}
\newcommand{\en}{\end{equation}}
\newcommand{\bea}{\begin{eqnarray}}
\newcommand{\ea}{\end{eqnarray}}

\newcommand {\prd}[3] {{\it Phys.\ Rev.\ }{\bf D #1} (#2) #3}

\newcommand {\plb}[3] {{\it Phys.\ Lett.\ }{\bf B #1} (#2) #3}
\newcommand {\npb}[3] {{\it Nucl.\ Phys.\ }{\bf B #1} (#2) #3}
\newcommand {\prl}[3] {{\it Phys.\ Rev.\ Lett.\ }{\bf #1} (#2) #3}
\newcommand {\jhep}[3] {{\it J. High Energy Phys.\ }{\bf #1} (#2) #3}

\newcommand {\heplat}[1] {{\tt hep-lat/#1}}
\newcommand{\hepth}[1] {{\tt hep-th/#1}}

\begin{document}

\section{Introduction}

The effective string 
description of the quark confinement 
predicts measurable effects on physical observables of the 
gauge theory, produced by the quantum fluctuations of that 
string \cite{lsw}. The most widely known is the L\"uscher  
correction to the inter-quark confining potential $V$ at 
large distance $r$
\eq
V(r)=\sigma\,r+2\,\mu-(d-2)\frac\pi{24r}+O(1/r^2)\,,
\label{pot}
\en
where $\sigma$ is the string tension  and $\mu$ a self-energy term. The 
attractive, universal Coulomb-like correction is known as the L\"uscher term.

A similar universal effect has been found in the low temperature behaviour 
of the string tension \cite{pa}
\eq
\sigma(T)=\sigma-(d-2)\frac\pi6 T^2+O(T^4)\,.
\label{st}
\en
Both effects may be summarised by saying that the infrared limit of the 
effective string is described by a two-dimensional conformal field 
theory (CFT) with conformal anomaly $c=d-2$. In this language, the 
(generalised) L\"uscher term $-\frac{c\,\pi}{24\,r}$ is the zero-point energy 
of a 2D system of size $r$ with Dirichlet boundary conditions, while the 
$-\frac{c\,\pi}{6}T^2$ term 
is the zero-point energy density in a 
very long cylinder (i.e. the string world-sheet of the Polyakov correlator) 
of period $L=1/T$ \cite{bcn}. 

We  proposed a method \cite{paper} to extend these 
results to a more general class of confining objects of $\sun$ gauge theories, 
the k-strings, describing the infrared behaviour of the flux tube joining 
sources made with $k$ fundamental representations.      

The spectrum of k-string tensions has been extensively studied in 
recent years, in the continuum \cite{ds,bary,jr} as well  as           
on the lattice \cite{lt1,ddpv}. 
So far, in numerical analyses one typically measured  the 
temperature-dependent k-string tensions $\sigma_k(T)$ through 
the Polyakov correlators and then extrapolated to $T=0$ using (\ref{st}), 
hence assuming  a  free bosonic string behaviour. 
  
 From a theoretical point of view there are good reasons
to expect values of $c$ larger than $d-2$. Indeed the conformal anomaly 
can be thought of as counting the number of massless degrees of freedom of the 
k-string. Therefore the relevant degrees of freedom are not only the 
transverse displacements but also the splitting of the k-string into its 
constituent strings. If the mutual interactions were negligible,  
each constituent string could vibrate independently so we had 
$c=k\,(d-2)$. Thus we expect that $c$ can vary in the range
\eq
d-2 \le\, c\,\le k\,(d-2)\,. 
\en
We studied this question in a class of 3D $\Z_4$ gauge models which admit a 
stable 2-string. At variance with the $\sun$ models considered so far, 
our model
depends on two coupling constants $\alpha$ and $\beta$. As a consequence, 
the continuum limit of the string tension ratio is not a constant, 
but a function $\sigma_2/\sigma=f(\lambda\equiv\alpha/\beta$) varying 
continuously from $\lambda=0$, where the system reduces to a pair of 
decoupled $\Z_2$ 
gauge systems, hence $\sigma_2/\sigma=2$, to $\lambda=1$, where it  
becomes the gauge dual of the 4-state Potts model with $\sigma_2/\sigma=1$.
Correspondingly also the central charge    is an unknown function
$c=c(\lambda)$  with $c(0)=2$ and $c(1)=1$. Our simulations were made at two
different values of $\lambda$ in the interval (0,1). There we studied 
the Polyakov-Polyakov correlators in the fundamental $(f)$ and double 
fundamental $(f\otimes f) $ representations at temperatures in the range 
$T\simeq T_c/2$ 
where experience on other gauge systems 
indicates that  the first universal temperature-dependent 
corrections show up and the expected functional form of this 
correlator is the Nambu-Goto two-loop expression is 
\eq
\bra P(0) P^{\dagger}(R)\ket_{T=1/L}\propto
\frac{e^{-2\mu\,L-\sigma\,RL-\frac{\pi^2[2E_4(\tau)-E_2^2(\tau)]}
{1152\sigma\,R^3}}}{\eta(\tau)}\,\,,
\label{ng}
\en  
where $\eta$ is the Dedekind eta function , $E_j$  denote the Eisenstein 
functions and $\tau=\frac{iL}{2R}$ is the aspect ratio of the cylinder 
associated to the string world-sheet. While the Polyakov correlators in 
the fundamental representations fitted very well this formula with  
stable parameters in a wide range of $R$, the similar fit with the 
correlator  the $f\otimes f$ sources 
turns out to be rather poor and the Ansatz  (\ref{ng}) does not result 
in stable parameters  \cite{z4kstr_pietro}.  

On the other hand, as explained in the next Section, there is a simple 
scaling argument suggesting a precise value for the central charge of 
the k-string.

\section{Scaling form of the Polyakov correlators}
\label{scaling}
The string picture of confinement fixes, in the IR limit, the 
functional form of the vacuum expectation value of gauge invariant 
operators. For instance the asymptotic behaviour in the region 
$2R\gg L$ of the correlation function  (\ref{ng}) is        
\eq
\bra P_f(0)\,P^\dagger_f(R)\ket_T \propto
\frac1{\sqrt{R}}\exp\left[-\sigma(T)\,R\,L-2\mu\,L\right]\, .
\label{ppst}
\en

Similar expansions are expected to be valid also for Polyakov correlators 
describing more specific features of $\sun$ gauge theory, like 
those involving baryonic vertices.

A baryon vertex is a gauge-invariant coupling of $N$ multiplets in the 
fundamental representation $f$ which gives rise to
configurations of finite energy, or baryonic potential, with $N$ 
external sources. 
At finite temperature $T$ these 
sources are the Polyakov lines $P_f(\vec{r}_i)$.

In the IR limit at finite temperature, i.e. 
$\vert\vec{r}_i-\vec{r}_j\vert\gg L ~ \forall i\not=j$,
we assume, in analogy with (\ref{ppst}) and the $N=3$ case,
\eq
\bra\, P_f(\vec{r}_1)\,P_f(\vec{r}_2)\dots P_f(\vec{r}_N)\,\ket_T
=e^{-F_N}
\sim\exp\left[-\sigma(T)\,L\,\ell(\vec{r}_1,\vec{r}_2 ,\dots,\vec{r}_N)
-N\mu\,L\right]\,,
\label{pnst}
\en
or, more explicitly,
\eq
F_N(\ell,L)\sim\sigma\,L\ell-(d-2)\frac{\pi\,\ell}{6\,L}+N\mu\,L~, \,
~(\vert\vec{r}_i-\vec{r}_j\vert\gg L ~ \forall i\not=j)\,,
\label{fb}
\en
where the coefficient of the $\ell/L$ term specifies that in this 
IR limit the behaviour of the baryon flux distribution is described by a 
CFT with conformal anomaly $d-2$  on the string world-sheet singled out 
by the position of the external sources. 

When $N>3$, depending on the positions $\vec{r}_i$ of the 
sources, some fundamental  strings of the baryon vertex may coalesce into 
k-strings \cite{Hartnoll:2004yr,bary}. As a consequence, the shape of the 
world-sheet changes in order to balance the string tensions and $\ell$ becomes 
a weighted sum, where a k-string of length $\lambda$ contributes with a term
$\lambda\,\frac{\sigma_k(T)}{\sigma(T)}\,$, with $\sigma(T)$ given by 
(\ref{st}), while
\eq
\sigma_k(T)=\sigma_k-c_k\,\frac\pi6 T^2+O(T^3)\,,
\label{skt}
\en
where $c_k$ is the conformal anomaly of the k-string. 

\begin{figure}[ht]
\begin{minipage}{70mm}
\begin{center}
\includegraphics[width=6cm,bb=115 315 515 545,clip]{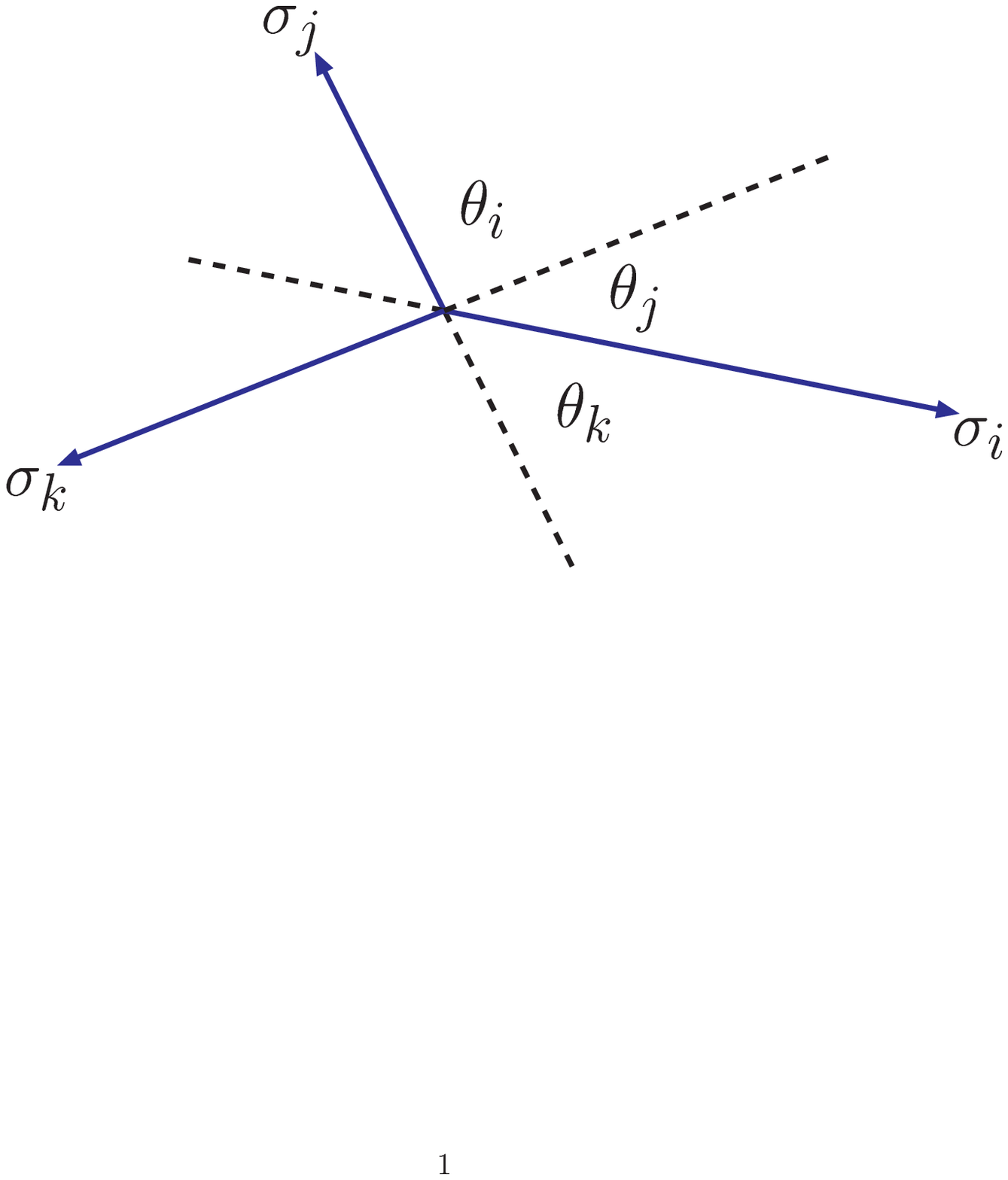}
\end{center}
\end{minipage}
\begin{minipage}{70mm}
\begin{center}
\includegraphics[width=4cm,bb=285 375 495 490,clip]{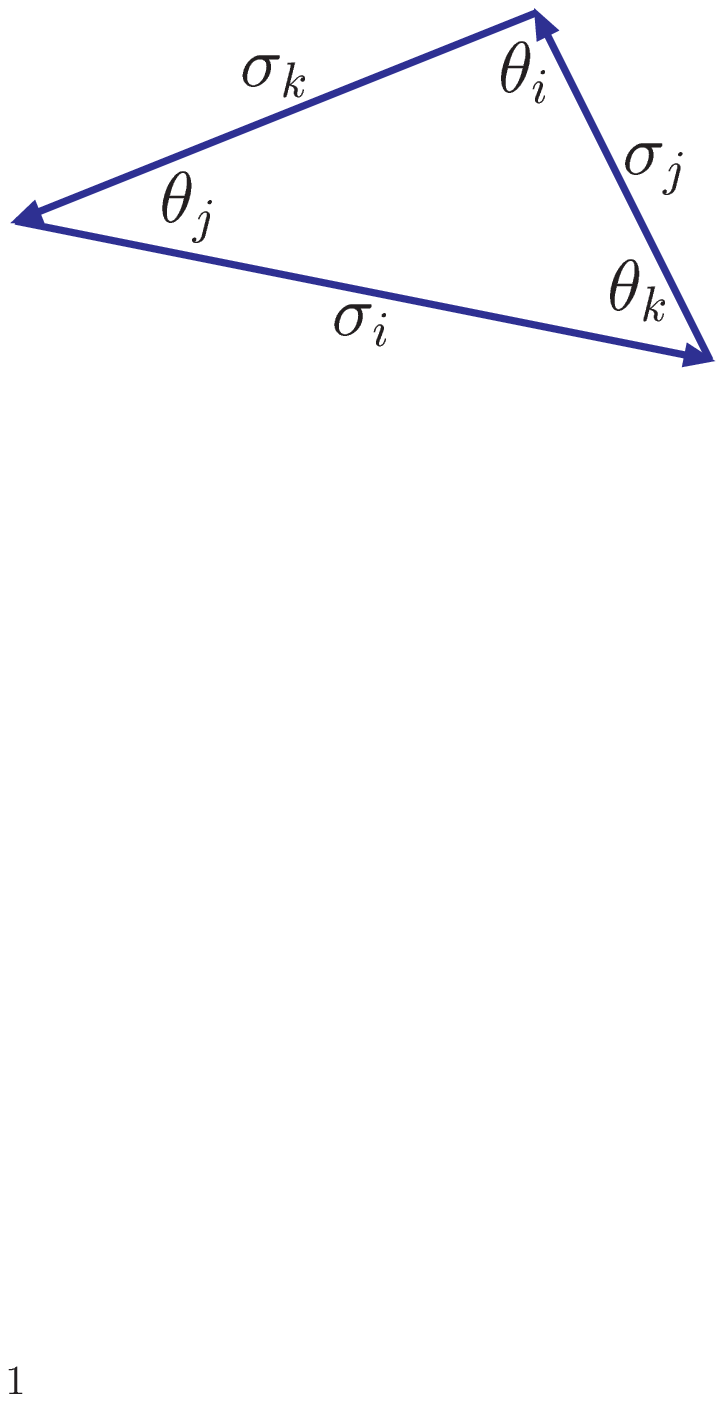}
\end{center}
\end{minipage}
\caption{The balance of the string tensions.}
\label{Figure:3}
\end{figure}

More generally, the baryonic free energy keeps the expected asymptotic 
form (\ref{fb}) only if the world-sheet shape does not change while varying 
$T$. Now  in a generic string configuration contributing to $\sun$ baryon  
potential, the angles at a junction of three arbitrary k-strings 
are given by (see Fig.\ref{Figure:3}, left)
\eq
\cos\theta_i=\frac{\sigma_j(T)^2+\sigma_k(T)^2-\sigma_i(T)^2}
{2\,\sigma_j(T)\,\sigma_k(T)}
\label{eqangles}
\en
and  others obtained by cyclic permutations of the indices 
$i,j,k$. 
Because by changing the temperature we can have only similar triangles 
(Fig.\ref{Figure:3}, right), all the string tension ratios are constant.
Moreover, considering an expansion of Eq.~\ref{eqangles} 
up to $T^3$ terms, we can connect it to the conformal anomaly:
\eq
\frac{c_i}{\sigma_i}=\frac{c_j}{\sigma_j}=\frac{c_k}{\sigma_k}=
\frac{(d-2)}{\sigma} \, ,
\en 
which leads directly to 
\eq
\sigma_k(T)=\sigma_k-(d-2)\frac{\pi\,\sigma_k}{6\,\sigma}\,T^2+O(T^3) \, ,
\label{main}
\en 
this being the main result of our work.

\section{Measurement of the conformal anomaly}
\label{conformal}
It is widely believed that the IR properties of the confining string are
described by a 2D CFT. This means that when the size of the string world-sheet 
is much larger than any other relevant length scale,  there are, besides the 
usual area and perimeter contributions, universal shape effects 
which are controlled by the conformal anomaly $c$ of the theory. However, 
there is no simple, general recipe to describe the way $c$ enters into 
these shape effects. 

For instance, if we denote with $F_{cyl}(\tau\equiv i\frac L{2R})$ 
the free energy of the CFT describing the IR behaviour of the k-string 
coupled to the representation $\R$,
provided that \emph{both} $R$ and $L$ are larger than $1/T_c$, which is the 
minimal distance at which the confining string picture applies,  we can 
write the Polyakov correlator 
in the form
\eq
\bra P_\R(0)\,P_\R^{\dagger}(R)\ket_{T=1/L}=
e^{-2\mu\,L-\sigma_\R\,RL-F_{cyl}(\tau)+O(1/\sigma_\R RL)}~~,
\en    
where the functional form form $F_{cyl}$, as anticipated in the Introduction, 
is known only in the two 
limits \cite{bcn}
\bea
F_{cyl}(\tau)=-\pi\frac LR\frac{c-24\,h}{24}&~~~~,~~L\gg 2R \, ; \\
F_{cyl}(\tau)=-\pi\frac RL\frac{c}{6}&~~~~,~~L\ll 2R \, ;
\label{ccc}
\ea 
$h$ being the scaling dimension of the lowest physical state which 
can propagate along the periodic direction. Outside these limits, 
the functional form depends explicitly on the whole spectrum of the CFT. 
It is know that these universal terms have the same origin of 
the quantum Casimir effect in 3D  and share with it the extreme 
sensitivity to the shape of the system.
Thus  any deformation of 
the boundary of the cylinder, like those considered for instance in the 
smearing procedure, produce modifications of the functional form which are 
difficult to calculate and to control. 

For these reasons the standard method generally 
employed to measure the conformal anomaly (the one we adopt here)
is  based solely on the infrared limit (\ref{ccc}) of a regular cylinder.

In practise we consider Polyakov correlators associated  to very long 
cylinders in the region $2R\gg L\gg 1/T_c$ and look after the stability of 
the fit to (\ref{ccc}).

\section{Numerical test}

The  general scaling argument developed in 
Section \ref{scaling} on the finite temperature corrections 
of the k-string tensions suggests a different behaviour with 
respect to the usual assumption that these corrections are those 
produced by a free bosonic string.
Since the comparison with theoretical predictions of k-string 
tensions is sensitive to this behaviour, it is important to 
check its validity.

We addressed such a question with a lattice 
calculation in a 3D $\Z_4$ gauge theory which is perhaps 
the simplest gauge system where there is a stable 2-string. 

Exploiting Kramers-Wannier duality and a suitable flips 
of the couplings of the corresponding spin model, 
we can insert any Polyakov correlator 
directly in the Boltzmann factor \cite{Giudice:2006hw},   
producing results with very high precision. 
Because of this last property it is possible to measure in a single 
numerical experiment the ratio
of the correlation of two Polyakov lines in the two representations: 
$G(r)_{ff}/G(r)_{f}$.

We performed  our Monte Carlo simulations on the AT model, at two different
points of the confining region, for which we previously measured the 
string tensions \cite{z4kstr_pietro}: ($\alpha=0.05,\beta=0.207$) and 
($\alpha=0.07,\beta=0.1975$). We worked on a  cubic 
periodic lattice of size $128\times128\times N_\tau$ with $N_\tau$ chosen 
in such a way that temperature of our simulations ranged from $T/T_c\simeq0.5$
to $T/T_c\simeq0.8$ and we took the averages over $10^6$ configurations in 
each point.

\FIGURE{
\includegraphics[width=6 cm]{ckfig/graf-corr-tot.eps}
\caption{Plot of $G(r)_{ff}/G(r)_{f}$ on a log scale.}
\label{Figure:ratiocorr}
}

The large distance behaviour of the data is well described by  a 
purely exponential behaviour (see Fig.\ref{Figure:ratiocorr})

\eq 
G(r)_{ff}/G(r)_{f}=\frac{\bra\,P_{ff}(\vec{r}_1)\,P_{ff}(\vec{r}_2)\ket_T}
{\bra\,P_f(\vec{r}_1)\,
P^\dagger_f(\vec{r}_2)\ket_T}\propto e^{-\Delta\sigma\,r\,N_\tau}\,,
\label{expo}
\en
with $\Delta\,\sigma=\sigma_{ff}-\sigma_f$.  
It is important to note that the logarithmic term, which is a 
potential source of systematic 
errors when neglected in Polyakov correlators, here is cancelled in the ratio.
Since (\ref{expo}) is an asymptotic expression, valid in the IR limit, we 
fitted the data to the exponential by progressively discarding the short 
distance points and taking all the values in the range 
$r_{min}\leq r\leq r_{max}=60\,a$ with $r_{min}$ varying from 15 to 40 lattice
spacings $a$. The resulting value of $\Delta\sigma$ turns out to be 
very stable. 
The whole set of  values of 
$\Delta\,\sigma(T)$ as functions of the inverse temperature $N_\tau=1/T$ are 
reported in Table 1 in Ref. \cite{paper}.

\begin{figure}[ht]
\begin{minipage}{70mm}
\begin{center}
\includegraphics[angle=270,width=6 cm]{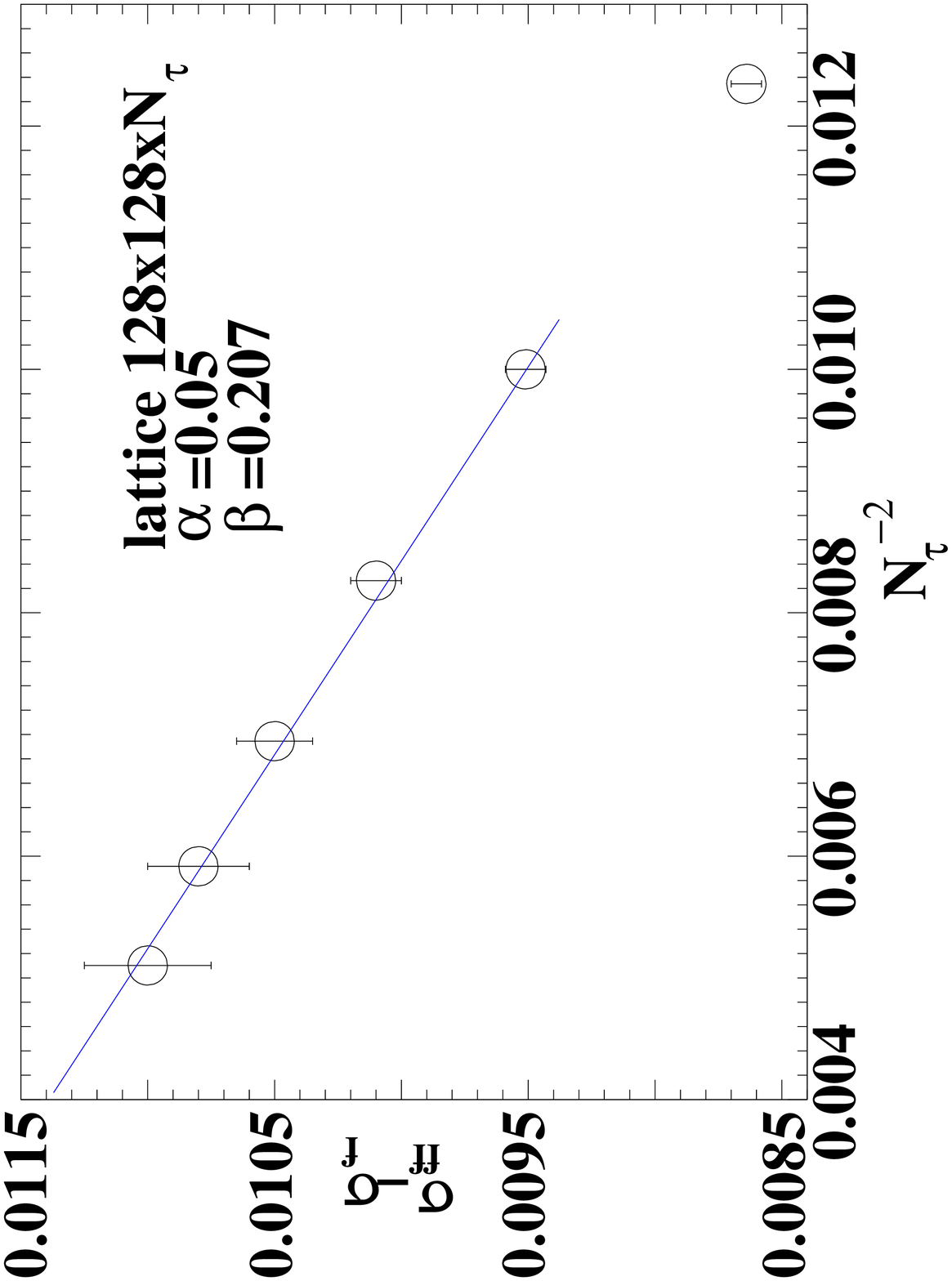}
\end{center}
\end{minipage}
\begin{minipage}{70mm}
\begin{center}
\includegraphics[angle=270,width=6 cm]{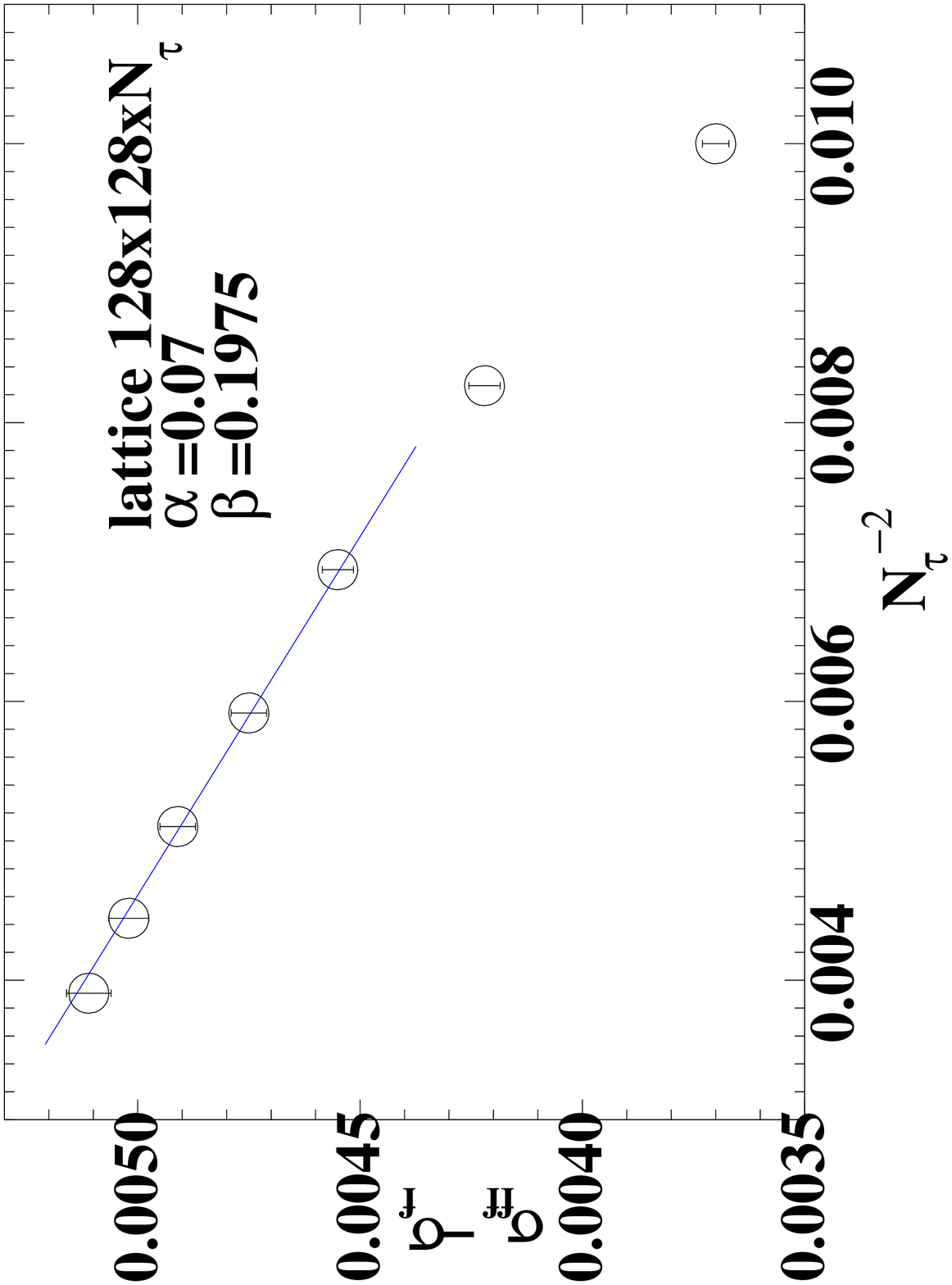}
\end{center}
\end{minipage}
\caption{Values of $\Delta\sigma(T)$  versus $T^2=1/N^2_\tau$ 
for two points of the parameter space. The solid line is 
a one-parameter fit to Eq.(4.2).}
\label{Figure:6}
\end{figure}

According to Eq.(\ref{main}), in the low temperature limit we expect
the asymptotic behaviour
\eq
\Delta\,\sigma(T)=\Delta\,\sigma(0)\left(1-\frac\pi{6\,\sigma}T^2\right)
     + O(T^3) \,.
\label{fit}
\en
Assuming for $\sigma$ the values estimated in \cite{z4kstr_pietro},
we used $\Delta\sigma(0)$ as the only fitting parameter. Neglecting one or 
two points too close to $T_c$, we got very good fits to (\ref{fit}) 
as shown in Figs. \ref{Figure:6}.
It is important to note that $\Delta\,\sigma(0)$ 
agrees with the difference of the previous estimates  \cite{z4kstr_pietro},
however the error is reduced by a factor of 25 
(first set of data) and even of 50 (second set of data).
The reason of this gain in precision is due to the fact that 
$\sigma_2$ was evaluated from a fit to (\ref{ppst}), even taking in account 
the Next-to-Leading-Order terms, which was rather poor 
because the 2-string does not behave as a free bosonic string 
\cite{z4kstr_pietro}. On the contrary our fits to (\ref{expo}) and 
(\ref{fit}) are very stable and the corresponding reduced $\chi^2/d.o.f$ 
are of the order of 1 or less.   

\section{Discussion}
We pointed out that a  simple argument on scaling properties 
of baryonic vertices suggested that the central charge of the CFT 
describing the infrared limit of a k-string is 
$c_k=(d-2)\frac{\sigma_k}\sigma$.

We checked it for $k=2$ in two different points 
of the phase diagram of 3D $\Z_4$ gauge model, where
the string tension ratio can be adjusted continuously in the interval 
$1\le   \frac{\sigma_k}\sigma\le 2$. In order to measure directly the 
central charge we used   
long cylindric string world-sheets  with aspect ratios $2R/L\gg 1$, where  
$R$  is the length of the open string 
(i.e. the distance between the Polyakov loops) and $L$  the length 
of the closed string. The fits involved two fitting parameter and
were particularly stable: when progressively discarding short 
distance data the fitted parameters showed always wide plateaux of more 
than 20 lattice spacings, with a $\chi^2/d.o.f$ of order 1 or less for   
more than 40 degrees of freedom for each temperature considered.  
 
In a  related work presented at this conference \cite{Bringoltz:2007ma}
 the k-string tensions
of some 3D $\sun$ gauge systems have been studied with a different method 
and the conclusion is different in the sense that a good agreement with 
the free bosonic string has been reported. Of course there is no compelling 
reason  for a 2-string of the $\Z_4$ system  to behave as the 2-string 
of a $\sun$ system, however it is possible that the discrepancy in the 
conclusions is due to the different methodology adopted. 
In \cite{Bringoltz:2007ma}
 the whole information is extracted from the correlation matrix of vanishing 
transverse momentum, smeared Polyakov loops. These   loops 
are much longer than ours, however their  mutual minimal distance considered
  in the zero momentum projection, hence 
the maximal contribution in the correlation matrix, is  few 
lattice spacings. In other terms, the 
 world-sheets which  maximally contribute to the correlation matrix are 
short cylinders with  very small aspect ratios
$2R/L\ll 1$ which is just the opposite of the limit where the central charge 
shows up as indicated in (\ref{ccc}). It is generally believed that in 
the process of diagonalisation of the correlation matrix, to get 
the projection on the ground state, the  non-universal contributions 
which  dominate at short distance are washed out, however it is 
difficult to fully understand 
in the context of the underlying effective string theory how the random 
boundary deformations 
of short cylinders implied by the smearing process could encode the information
on the conformal anomaly, which is an essentially IR quantity 
that can be directly observed only in long cylinders.   
According to the general considerations made in Section \ref{conformal},
it would be highly desirable to develop a similar correlation matrix analysis
where the minimal loop distance taken into account in the zero momentum 
projection  is larger than $1/T_c$, 
where the effective string picture starts working.

\acknowledgments
We are grateful to Michele Caselle, Paolo Grinza and Ettore Vicari for useful
discussions.

\end{document}